\documentstyle[11pt,newpasp,twoside,epsfig]{article}
\def\I{{\'\i}}
\def\msun{M$_{\odot} \ $}

\def\edcomment#1{\iffalse\marginpar{\raggedright\sl#1\/}\else\relax\fi}
\reversemarginpar

\begin{document}
\title{ Enrichment of the Intergalactic Medium by the Milky Way Subgroup}
 \author{Leticia Carigi}
\affil{Instituto de Astronom\I a, UNAM, Apdo. Postal 70-264, CP 04510, M\'exico DF, Mexico}

\begin{abstract}
I study the effect of the galaxies of the Milky Way subgroup on the chemical enrichment of
the diffuse gas within the Local Group.
The preferred model predicts that the present-day intergalactic medium mass is 1/22 of the initial
baryonic mass and its metallicity is 1/20 $Z_{\odot}$.
Nearly 80 \% of intergalactic metals were produced in the Galactic bulge.
\end{abstract}

\section{Introduction}
The Local Group is made up of four subgroups:
the Milky Way, M31, the Local Group Cloud, and NGC 3109.

In this paper I will focus the attention only on the Milky Way subgroup (MWS).
This subgroup consists of our galaxy and 13 satellite galaxies:
two irregular galaxies (Small and Large Magellanic Clouds), one dwarf irregular galaxy (Leo A),
and 10 dwarf spheroidal galaxies (Sculptor, Phoenix, Fornax, Carina, Leo I, Sextans,
Leo II, Ursa Minor, Draco, Sagittarius).

The properties of the intergalactic medium (IGM) of the Local Group
are unknown, and its density
may be negligible.
In this work, I predict chemical properties for the near IGM,
based on chemical evolution models computed for the Milky Way and
its neighboring galaxies.

\section{ Chemical Evolution Models}

\subsection{Galactic Assumptions}

Every chemical evolution model of those galaxies that form the MWS has an age of 13 Gyr
and includes the following common assumptions:

i) Galaxies are formed by gradual gas accretion
of primordial material, 
which rate decreases exponentially
in time with an e-folding constant ($\tau$), which is different for each galaxy.

ii) The adopted IMF (Kroupa, Tout \& Gilmore 1993) covers the mass range  between 0.01 and 120 $M_{\odot}$.

iii) Yields by Maeder (1992) and Renzini \& Voli (1981) are used.

\noindent
The star formation rates and outflows are not different among the models and they will be explained 
in detail for each type of galaxies.

Since mass loss through galactic winds is the physical process considered in the
chemical enrichment of the IGM gas, in this section a general
review of the models for the galaxies forming the MWS will be given, making emphasis
on those galaxies that experiment galactic winds.  

Galactic winds are powered by supernovae of Types I and II:
These SNe heat the interstellar medium and when the thermal energy in the gas is
higher than the potential energy, the gas is blown away and the star formation stops.

\smallskip
The {\bf Milky Way galaxy} is modeled by 3 components:
halo, bulge and disk. 

The chemical evolution model for the {\it Galactic disk} is taken from Carigi (2000).
The {\it Halo} collapses very quickly, with  $\tau=0.01$ Gyr, 
reaching 2 $\times \ 10^9$ \msun in 0.7 Gyr.
Most of the models of  the halo and the disk do not predict or assume
the presence of galactic winds.

The {\it Galactic bulge} model is similar to that of  Moll\'a, Ferrini, \& Gozzi  (2000). 
The initial baryonic and  dark matter masses inside the bulge radius 
(2 kpc)
are 2 $\times \ 10^{10}$ \msun and  2 $\times \ 10^9$ $M_{\odot}$, respectively.
The bulge is built quickly with $\tau$= 1 Gyr.
The star formation rate is proportional to the gas mass, with an efficiency factor equal to 20 Gyr$^{-1}$.

\smallskip
The {\bf irregular galaxies}  are modeled by the closed-box approximation 
(or $\tau$ very short $\sim$ 0.001 Gyr). 
The star formation rate is proportional to the gas mass.
Carigi, Col\I n, \& Peimbert  (1999) and Larsen, Sommer-Larsen, \& Pagel (2001)
have found that O enriched winds are ruled out 
for this type of galaxies. Carigi et al. (1995) also discard galactic winds because they predict 
a high amount of
ejected mass, which is not detected. Nevertheless, Larsen et al. (2001) obtained that some dIrr can
be explained by galactic winds.

\smallskip
Four of the Local {\bf dwarf spheroidal galaxies} (Carina, Leo I, Leo II, and Ursa Minor) 
are modeled considering the 
star formation rates from Hern\'andez, Gilmore, \& Valls-Gabaud (2000).
This SFR is characterized by a series of smooth bursts.
Based on the star formation history, these galaxies are
classified in two types: single burst (Leo II and Ursa Minor) and two main bursts (Carina and Leo I).
A secondary and constant infall have been assumed only for dSph galaxies with double burst.

Carigi, Hern\'andez, \& Gilmore (2001) computed the amount and abundances of gas expelled by galactic winds
in these four dSphs. We have assumed that a galactic wind occurs, at a time $t_{wind}$,
 immediately after the maximum of a burst.

In the literature there are no chemical evolution models for the rest of the Local dSphs, 
nor star formation histories to make models. 
For that reason, the wind masses and chemical abundances
of Sculptor, Phoenix, Fornax, Sextans, Draco, and Sagittarius
were obtained by linear interpolation and extrapolation, according to the dynamical mass, of the values
 computed by Carigi et al. (2001).
For these 6 galaxies, the time of each galactic wind was estimated using 
the relative SFR showed by Mateo (1998),
for $t_{wind}$ I adopted the age indicated by main sequence stars.
Again, these galaxies were classified as
single or double burst.
Mateo (1998) and Carigi et al. (2001) have considered a galactic age of 15 Gyr, 
but in this work the $t_{wind}s$ have been adjusted to a galactic age of 13 Gyr.

\smallskip

Based on these chemical evolution models for galaxies,
I conclude that {\bf winds are important in:  the Galactic bulge and 
dwarf spheroidal galaxies } 

\begin{figure}
\hspace*{3.35cm}
\epsfig{figure=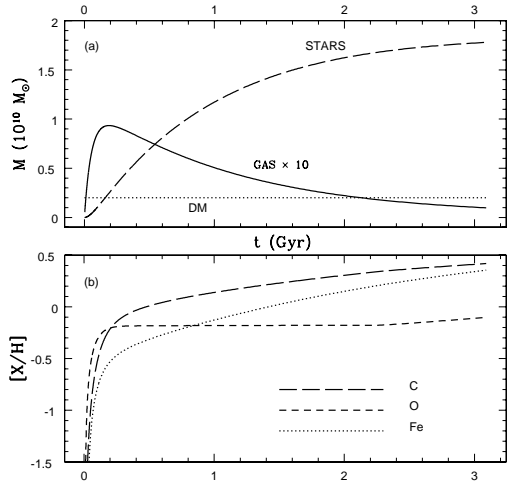,width=7cm}
\vspace*{0.00cm}
\caption{Predictions of the Galactic bulge model before the galactic wind occurs.
a) Dark matter, gas, and stellar masses as a function of time.
b) Time evolution of the abundances ratios.}
\end{figure}

\subsection{Intergalactic Assumptions}

I assumed that the MWS is built from an 
initial baryonic mass $ M_B(0)=8\times 10^{10} M_\odot$, which is higher than the sum of the luminous 
masses of the galaxies that form
this subgroup, that amounts to $7.7 \times 10^{10} M_\odot$.

The baryonic gas mass evolves as
\begin{equation}
\dot{M}_B(t) = W(t) - A(t),
\end{equation}
where $A$ is the accretion rate that formed the galaxies and 
$W$ is the rate of returned material to the IGM from
galaxies through galactic winds, $W(t)= \sum_i M_{wind}^i \delta(t -t_{wind}^i)$.
Infall stops when the galactic wind occurs.
I do not consider mergers of galaxies or transference of material between galaxies.

The $j$-element abundance of intergalactic gas, $X_j$, evolves as
\begin{equation}
X_j(t) M_B(t) = - \int_0^t X^A_j(t') A(t')dt' + \int_0^t X^W_j(t') W(t')dt' + M_B(0) X_j(0),
\end{equation}
where $X^A_j$, and $X^W_j$ are the $j$-element abundances in the  accreted gas
and the galactic winds, respectively.
I assume primordial values for the intergalactic gas initial abundances.
The infall abundance should be equal to $X_j$, but in this work I assumed it to be primordial.
This fact may affect the chemical evolution of slow-formation galaxies,
like the Galactic disk or in the second episode of formation of dSphs.
The predicted metallicity in these galaxies would be slightly higher, while
the IGM metallicity slightly lower.

In order to relate the time in chemical evolution models of the galaxies with redshift ($z$),
I have assumed the  $\Lambda$CDM model
with $\Omega_m=0.3$, $\Omega_\Lambda=0.7$, and $h=0.65$.
The age of the universe in this model  is 14.5 Gyr, and since the age of the MWS galaxies is 13 Gyr, 
the galaxies started building 1.5 Gyr after the Big Bang.

\section{Results}
In this section I present the model results for the Galactic bulge, the dSph galaxies, and 
the chemical evolution for the intergalactic gas inside the MWS.

\smallskip
{\bf Galactic Bulge}

Figure 1  presents, before the galactic wind happens, the Galactic bulge evolution of:
a) gas, stellar, and dark matter masses; and
b) abundance ratios in the gas.
The predicted abundances  are in agreement  with the observed values for most of the stars
(Mc William \& Rich 1994).

The galactic wind occurs at $t_{wind}=3$ Gyr, transporting 1 $\times \ 10^8$ \msun of gas to the IGM,
with C, O, Fe, and $Z$ abundances equal to  7.2, 5.9, 2.7, and 24.2 $\times \ 10^{-3}$, 
respectively.
The $t_{wind}$ depends, mainly,  on the amount of non-baryonic mass inside the bulge radius. 
Results for another amount of dark matter  are presented in the intergalactic gas subsection.

\begin{figure}
\hspace*{3.35cm}
\epsfig{figure=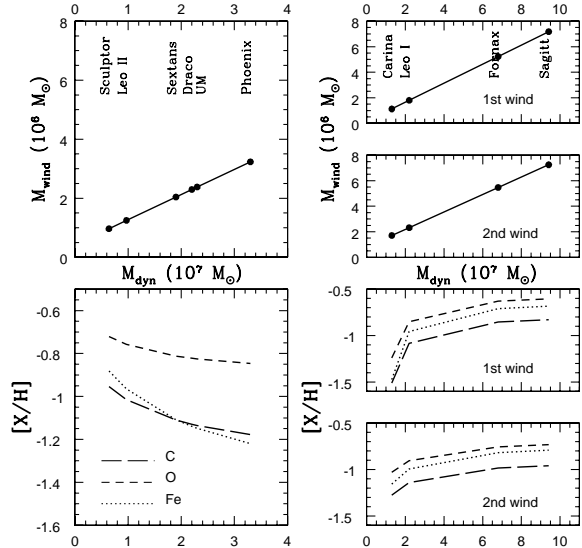,width=8cm}
\vspace*{0.00cm}
\caption{Predictions for MW dSphs.
Mass ejected by winds and abundance ratios versus dynamical mass.
}
\end{figure}

\smallskip
{\bf dSph Galaxies}

The mass ejected by galactic winds and its abundances predicted for all MW dSphs
are  plotted in Figure 2.
Carigi et al. (2001) predicted main galactic winds at 1.05, 5.01, 5.22, and 5.56 Gyr  
for Ursa Minor, Leo II, Carina, and Leo I, respectively;
and second winds at 10 Gyr for the last two dSphs. 
According to Mateo (1998)
Sculptor, Sextans, Draco and Phoenix had winds at 8.0, 11.0, 7.4, and 2.0 Gyr, respectively.
For Fornax and Sagittarius the first wind occurs at 2 and 3 Gyr, while  the second one occurs at 11.0 at 9.5 Gyr,
respectively.

\smallskip
{\bf Intergalactic Gas}

Figure 3 presents the evolution of the IGM mass,  the baryonic mass of all the galaxies, and 
the returned mass to the IGM via galactic winds.
One can see the IGM mass decreasing with time due to the formation of
the galaxies from this mass.
Each increase of the total wind mass is due to galactic winds of bulge and dSphs.
The important increases at $z \sim 1.5$ and $z \sim 0.5$ are caused by the bulge wind
and the secondary dSph winds, respectively.

\begin{figure}
\hspace*{3.35cm}
\epsfig{figure=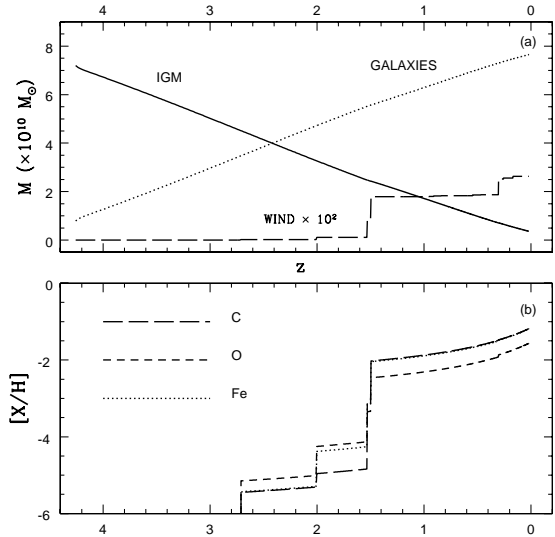,width=7.8cm}
\vspace*{0.00cm}
\caption{Predictions for the intergalactic medium.
a) IGM, galactic and ejected masses as a function of redshift.
b) Redshift evolution of the IGM abundance ratios}
\end{figure}

The  C, O, Fe, and $Z$ abundances of the intergalactic medium  are presented
in the same figure.
The predicted present-day metallicity is 0.05 $Z_\odot$, in agreement with the poorest high velocity clouds (HVC)
([M/H] $<$ -1 Sembach et al. 1999,  [Fe/H] = -1.5 Lu et al. 1998).
Abundances increase abruptly at $z \sim 1.5$, when the bulge wind takes place, 
and at that moment the proportion among elements is modified.
The bulge contributes with the 92, 65, 89, and 79 \% of the total C, O, Fe, and $Z$ mass
in the IGM.
C abundance is higher than O in the bulge, because $Z_{bulge}$ is supersolar at $t_{wind}$  and
most of the C in the bulge wind is produced by supermetallic stars and low-and-intermediate mass stars;
instead, the O is produced by massive stars with subsolar abundances.
At $t_{wind}=3$ Gyr SNIa have contributed to the Fe abundance.

Since the amount and metallicity of the gas between galaxies in the Local Group are unknown, 
models are not constrained, and therefore I computed other two models changing the main free parameter:
the initial baryonic mass in the MWS, in the first model, and the non-baryonic mass inside the bulge,
in the second model.

Blitz et al. (1998) suggested that if the HVC are extragalactic, then there is an important amount of
baryonic mass in HI around the Local Group, and this mass is similar to the disk mass.
For this reason, in the first extra model  I assumed that 
$ M_B(0)=8\times10^{10}+2\times10^{10}=10\times10^{10} M_{\odot}$.
This model predicts $Z_{IGM} \sim 0.01 Z_\odot$, five times lower than the observed $Z$ in the poorest HVCs,
result which implies that the HVC are not intergalactic or that the amount of baryonic mass in HI around the MW
is negligible.

In another experiment I assumed that in the Galactic bulge the dark matter mass is equal to the baryonic mass,
in this case the bulge does not have a wind and the predicted IGM metallicity is 0.01 $Z_\odot$.

\section{Conclusions}
From chemical evolution models of galaxies associated with the Milky Way subgroup, 
the following conclusions are reached for the intergalactic medium:
i) The present-day IGM mass is $\sim 0.05$ of the initial
baryonic mass, that is, $\sim 4 \times 10^9$ $M_{\odot}$, 
ii) The predicted current metallicity of the IGM is 0.05 $Z_{\odot}$,
iii) Nearly 80 \% of intergalactic metals come from the Galactic bulge.

I hope that these results motivate the search for the IGM in the Local Group.
The observed $Z$ of the IGM will constrain the models and might permit us to estimate
the  initial baryonic mass of the Local Group.

\acknowledgments
I am grateful to Vladimir Avila-Reese for illuminating discussions
and Manuel Peimbert for useful comments on the manuscript.


\begin{references}

\reference
Blitz, L., Spergel, D. N., Teuben, P.J., Hartmann, D., \& Burton, W. B. 1999,
\apj, 514, 818

\reference
Carigi, L. 2000, Rev. Mex. Astron. Astrof., 36, 171


\reference
Carigi, L., Col\I n, P., \& Peimbert, M. 1999,
\apj, 514, 787

\reference
Carigi, L., Col\I n, P., Peimbert, M., \& Sarmiento, A. 1995, \apj, 445, 98

\reference
Carigi, L., Hern\'andez, X., \&Gilmore, G., 2001,
\mnras, in preparation.

\reference
Hern\'andez, X.,  Gilmore, G., \&  Valls-Gabaud, D.: 2000,
\mnras, 317, 831.

\reference
Kroupa, P., Tout, C.A., \& Gilmore, G. 1993, \mnras, 262, 545

\reference
Larsen, T.I., Sommer-Larsen, J., \& Pagel, B.E.J. 2001,
\mnras, 323, 555

\reference
Lu, L., Sargent, W. L. W.,
 Savage, B. D., Wakker, B. P.;
 Sembach, K. R., \&  Oosterloo, T. A.
1998, \aj, 115, 162

\reference
Mateo, M. 1998, \araa,  36, 435

\reference
Mc William, A., \& Rich, R. M. 1994, \apjs, 91, 749

\reference
Moll\'a, M., Ferrini, F. \& Gozzi, G. 2000, \mnras, 316, 345
\araa

\reference
Maeder, A. 1992, \aj 264, 105 

\reference
Renzini, A., \& Voli, M. 1981, \aj, 94, 175

\reference
Sembach, K. R., Savage, B. D.,
 Lu, L., \& Murphy, E. M. 1999
\apj, 515, 108

\end{references}
\end{document}